\documentclass[11pt]{article}

\usepackage{float} 
\usepackage{amsmath}
\usepackage{amssymb}
\usepackage{array}
\usepackage{bbm}
\usepackage{makeidx}
\usepackage{hyperref}
\usepackage{cleveref}
\usepackage{caption}
\usepackage{subcaption}
\usepackage{amsthm}
\usepackage{color}
\usepackage{mathtools}
\usepackage{graphicx} 
\usepackage[LGR,T1]{fontenc}
\usepackage{mathrsfs}
\usepackage[noend]{algorithmic}
\usepackage[ruled,vlined]{algorithm2e}
\usepackage{lipsum}
\usepackage{booktabs}
\usepackage{xcolor}
\usepackage[title]{appendix}
\hypersetup{
    colorlinks,
    linkcolor={blue!80!black},
    citecolor={green!50!black},
}
\usepackage{apptools}
\usepackage{tikz}
\usetikzlibrary{arrows, positioning, automata}
\usepackage{comment}
\usepackage[shortlabels]{enumitem}

\usepackage[mathscr]{euscript}
\DeclareSymbolFont{rsfs}{U}{rsfs}{m}{n}
\DeclareSymbolFontAlphabet{\mathscrsfs}{rsfs}

\def\GS{{\sf GS}}

\def\cuC{\mathscrsfs{C}}
\def\cuE{\mathscrsfs{E}}
\def\cuU{\mathscrsfs{U}}
\def\cuL{\mathscrsfs{L}}

\def\de{{\rm d}}
\def\Par{{\sf P}}
\def\sTV{\mbox{\tiny \rm TV}}

\def\dist{{\sf dist}}

\def\<{\langle}
\def\>{\rangle}

\def\sE{{\sf E}}

\def\sTV{\mbox{\tiny \rm TV}}

\def\salg{\mbox{\tiny \rm alg}}

\def\prob{{\mathbb P}}

\usepackage{mathtools}

\newtheoremstyle{myremark} 
    {\topsep}                    
    {\topsep}                    
    {\rm}                        
    {}                           
    {\bf}                        
    {.}                          
    {.5em}                       
    {}  

\newtheorem{theorem}{Theorem}

\theoremstyle{myremark}
\newtheorem{remark}{Remark}[section]

\allowdisplaybreaks
\usepackage[top=1in,bottom=1in,left=1in,right=1in]{geometry}
\usepackage[utf8]{inputenc}

\def\bA{{\boldsymbol A}}

\def\cS{{\mathcal S}}

\def\cS{\mathcal{S}}

\def\bsigma{{\boldsymbol \sigma}}
\def\S{{\mathbb S}}

\def\bA{\mathbf{A}}

\def\bF{\mathbf{F}}
\def\bG{\mathbf{G}}

\def\bQ{\mathbf{Q}}

\def\bY{\mathbf{Y}}

\def\bg{\boldsymbol{g}}

\def\bm{\boldsymbol{m}}

\def\bw{\boldsymbol{w}}

\def\bA{\boldsymbol{A}}

\def\bF{\boldsymbol{F}}
\def\bG{\boldsymbol{G}}

\def\bQ{\boldsymbol{Q}}

\def\bY{\boldsymbol{Y}}

\def\normal{{\mathsf{N}}}

\def\bm{\boldsymbol{m}}

\def\conv{{\rm conv}}

\def\ALG{{\sf ALG}}

\def\reals{{\mathbb R}}

\def\bfzero{{\boldsymbol 0}}

\def\z{{\boldsymbol z}}
\def\u{{\boldsymbol u}}

\def\sP{{\sf P}}

\newcommand{\eps}{\varepsilon}

\newcommand{\RN}[1]{%
  \textup{\uppercase\expandafter{\romannumeral#1}}%
}

\newcommand{\RNum}[1]{\uppercase\expandafter{\romannumeral #1\relax}}



\makeatletter
\newcommand*{\rom}[1]{\expandafter\@slowromancap\romannumeral #1@}
\makeatother

\title{Optimization of random cost functions and statistical physics}

\author{Andrea Montanari
	\thanks{Department of Statistics and Department of Mathematics, Stanford University} 
}
\date{}
\begin{document}
\maketitle

\begin{abstract}
This is the text of my report presented at the 
 29th Solvay Conference on Physics on `The Structure and Dynamics of Disordered Systems'
 held in Bruxelles from October 19 to 21, 2023.
 
 I consider the problem of minimizing a random energy function $H(\bsigma)$, 
 where $\bsigma$ is an $N$-dimensional vector, in the high-dimensional regime
 $N\gg 1$. Using as a reference point a 1986 paper by Fu and Anderson, 
 I take stock of the progress on this question over the last 40 years. 
 In particular, I focus on the influence and ramifications of ideas
 originating from statistical physics.
 
 My own conclusion is that several of the most fundamental questions in this area (which
 in 1986 were barely formulated) have now received mathematically rigorous answers,
 at least in simple --yet highly nontrivial-- settings. Instrumental to this
 spectacular progress was the dialogue  between different research communities: physics, computer
 science, mathematics. 
\end{abstract}

\section{Introduction: Optimization in high dimension}

This report is concerned with the problem of optimizing an energy (or cost) function 
$H$ in a high-dimensional space:
\begin{align*}
\mbox{minimize} &  \;\;\;\; H(\bsigma)\, ,\\
\mbox{subj. to} &  \;\;\;\; \bsigma\in \cS^N = \cS\times \cdots \times \cS\, .
\end{align*} 
In statistical physics, $H$ is referred to as the `Hamiltonian,' and  the coordinates
$(\sigma_1,\dots,\sigma_N)=:\bsigma$ are called `spins'
or `spin variables.' Any global minimizer is called a `ground state.'
While this terminology is related to the physics of these models, 
its origin will not be important in what follows.
Typical examples of the space $\cS$ studied in the literature are
 $\cS=\{+1,-1\}$ (`Ising spins'), $\cS=\{1,2,\dots,q\}$ (`Potts spins'), $\cS  = \reals$,
 (`soft spins'),
and so on. Below we will also consider the case $\bsigma\in \S^{N-1}(\sqrt{N})$,
i.e., $\bsigma$ belonging to the sphere of radius $\sqrt{N}$ in $N$ dimensions (`spherical spins').
 This is not
a product space, but for many purposes behaves analogously.

Of course, as the dimension $N$ grows, the space of possible Hamiltonians $H$ 
explodes. For instance, if $\cS^N=\{+1,-1\}^N$, then specifying an Hamiltonian
requires, in general $2^N$ numbers. In such generality, the optimization problem is not
particularly interesting: finding a near optimizer requires exhaustive search over the $2^N$ 
configurations. Its complexity is linear in the input size (the number of parameters of the 
Hamiltonian).

In practice, we are interested in families of Hamiltonians,
with a special structure that allows to specify them with a number of parameters that is polynomial 
in $N$. Here are two examples (where $[N]:=\{1,2,\dots,N\}$ is the set of first $N$ integers, and 
$\binom{[N]}{k}$ is the set of its subsets of size $k$):
\begin{align*}
H_1(\bsigma)  = \sum_{R\in\binom{[N]}{k}}\psi_R(\bsigma_R)\, ,\;\; \bsigma_R := (\sigma_i:i\in R)\, ,
\;\;\;\;\;\;\;
H_2(\bsigma)  = \sum_{a=1}^m\psi(\<\bw_a,\bsigma\>)\, .
\end{align*}
In words, in the first example, $H_1$ is a sum of terms each involving at most
$k$ coordinates of $\bsigma$, and in the second example $H_2$ is a sum of terms each 
involving a one-dimensional projection.

\begin{figure}
\hspace{2cm}\includegraphics[width=6cm,angle=0]{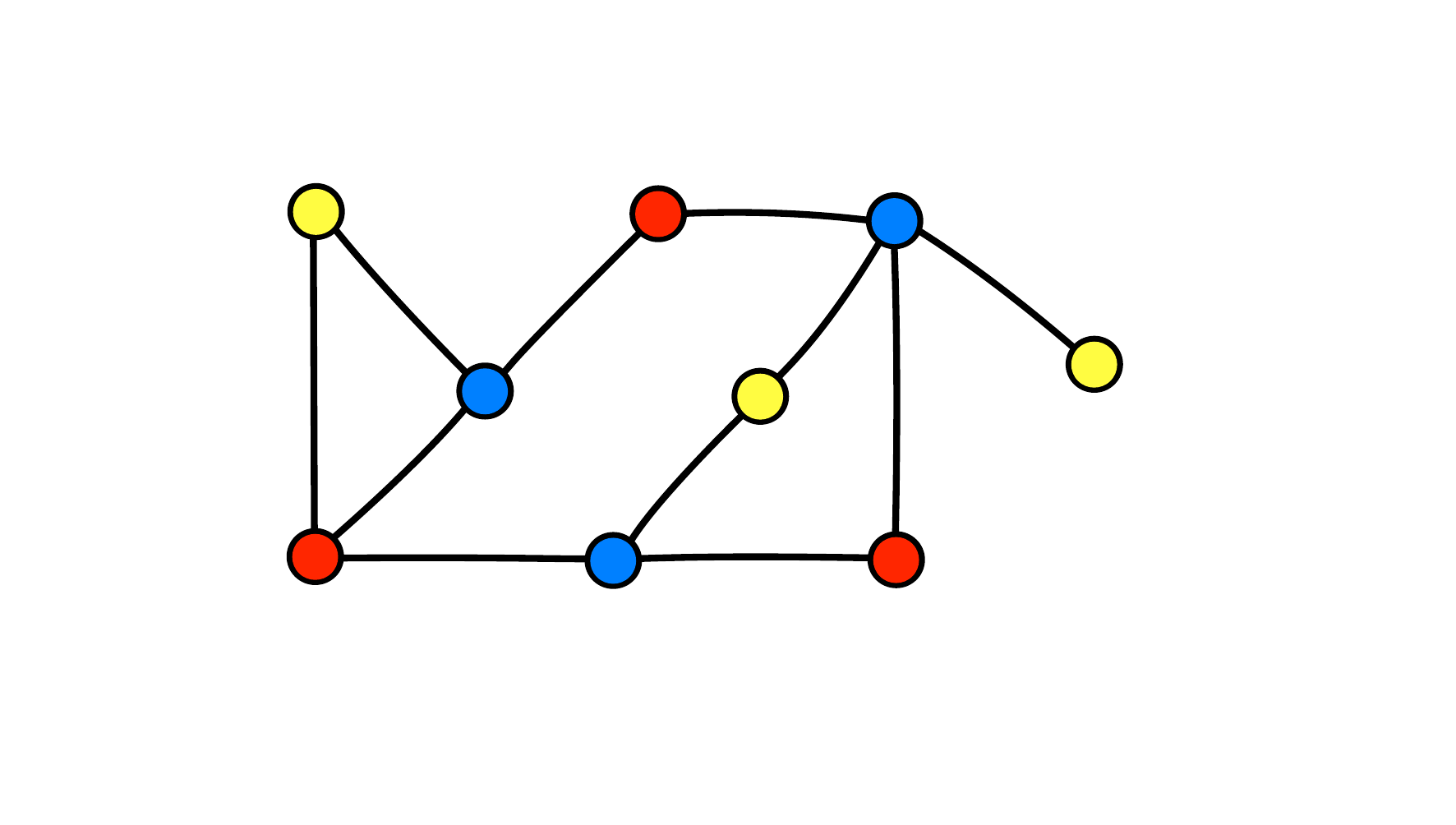}

\phantom{A}\vspace{-4cm}

\phantom{A}\hspace{8cm}\includegraphics[width=3cm,angle=-90]{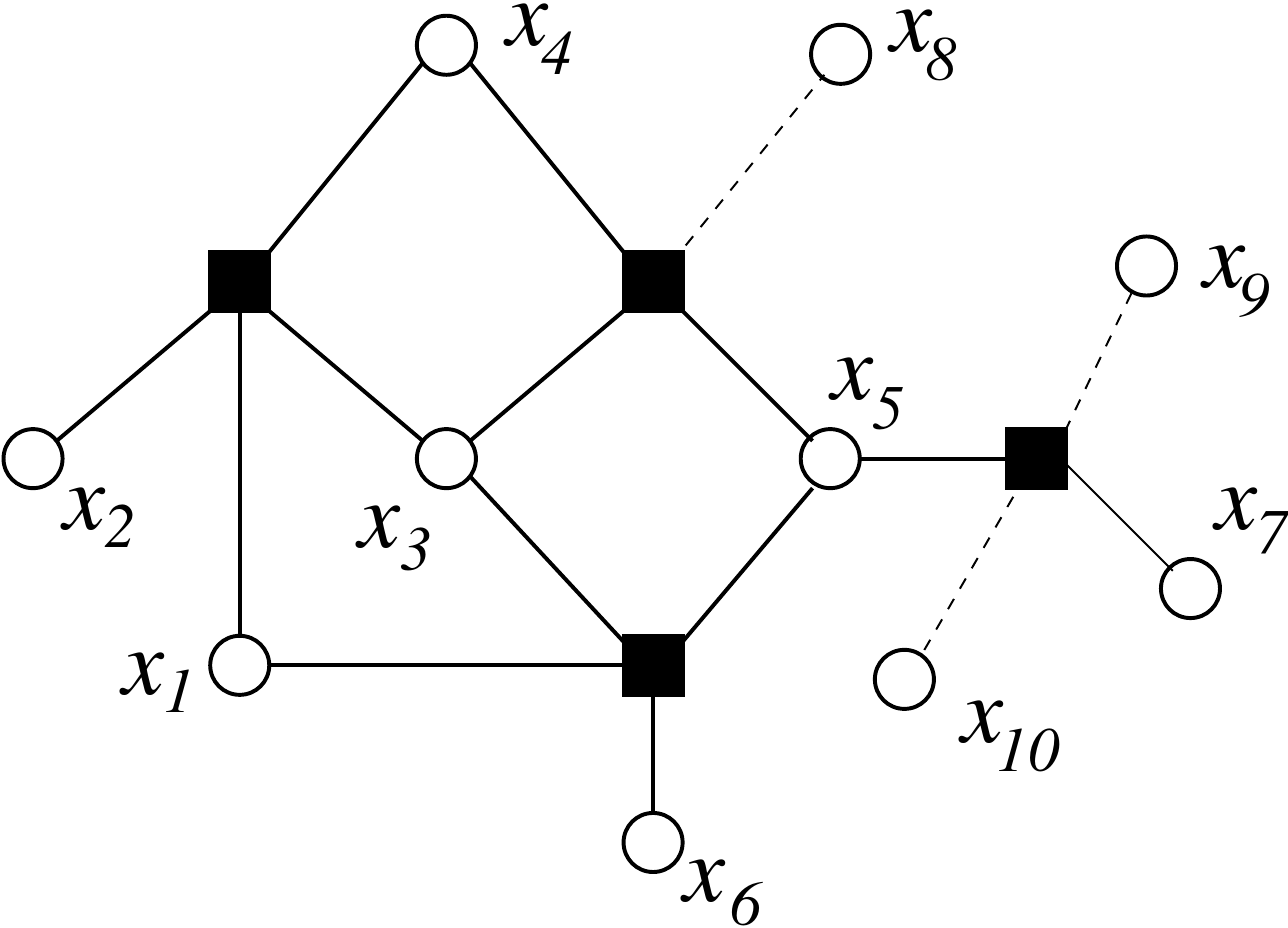}
\vspace{0cm}
\caption{Left: graphical representation of a graph-coloring problem. Right: factor graph representation
of a satisfiability problem.}\label{fig:Examples}
\end{figure}

A large number of optimization problems of classical interest in computer science can be 
written in this form, see Fig.~\ref{fig:Examples}. 
Examples include graph coloring, satisfiability, maxcut, graph partitioning, and so on.
We refer to \cite{MeMo} for background on these models. 
The same setting also includes optimization problems arising
in high-dimensional statistics and machine learning. In these cases $\bsigma$ is an unknown 
vector of parameters or a signal that we want to reconstruct (an image, an electromagnetic
 signal, the labels
of datapoints in a clustering problem) from some collected data $\bY$ (e.g. noisy linear measurements 
of the electromagnetic signal). One possible reconstruction method is to minimize the negative log-likelihood
 $H(\bsigma)= -\log{\rm P}(\bY|\bsigma)$ over a suitable parameters space.
 We refer to Lenka Zdeborov\'a's report
 in these proceedings for examples (cf. also \cite{montanari2022short}). 
 
 In many cases of interest (most notably in high-dimensional statistics and machine learning)
 the function $H$ is itself random, and we will denote probabilities with
 respect to this randomness by $\prob$. The general problem we face is, informally:
 \begin{quote}
 Given a realization of the energy function $H(\,\cdot\,)\sim \prob$ find  
 (in time polynomial in $N$) $\bsigma^{\salg}\in\cS^N$
 such that $H(\bsigma^{\salg}) \approx \min_{\bsigma\in\cS^N}H(\bsigma)$,
  with high probability
 (with respect to the random choice of $H$).
 \end{quote} 
 This report attempts an overall assessment of the progress on this problem
 over the last forty years, with a special focus on ideas originated from statistical 
 physics. For a somewhat more technical review, I refer to \cite{auffinger2023optimization}.

%
%
\subsection{Two starting points: Optimization theory and statistical physics}

In order to appreciate the extent to which interactions between physics and computer science 
have been crucial for the evolution of this subject, it is useful to evoke
some reference points in each of these two research areas. For clarity, we will refer to a concrete
energy function. Given a graph $G=(V=[N],E)$, define
\begin{align}
H(\bsigma) = \sum_{(i,j)\in E}\sigma_i\sigma_j\,, \;\;\;\;\; \bsigma\in\{+1,-1\}^N\, .
\label{eq:Hmaxcut}
\end{align}
In statistical physics terms, this is the energy function (or Hamiltonian) 
of an antiferromagnetic Ising model. 
Minimizing $H$ corresponds to finding a maximum cut in $G$.

A generalization of the above Hamiltonian is given by the quadratic function (pairwise Ising model):
\begin{align}
 H(\bsigma) = \<\bsigma,\bA\bsigma\>\, , \;\;\;\;\; \bsigma\in\{+1,-1\}^N\, .
 \label{eq:H-Ising}
\end{align}
The corresponding minimization problem is sometimes referred to as the `little Grothendieck problem'
\cite{alon2006approximating}.

Let us emphasize two important elements of the standard computer science approach.
\begin{enumerate}
\item \emph{Adversarial model.} Instances of a given optimization problem are thought 
as chosen by an adversary whose objective is to make our algorithm perform as poorly as possible.
For instance, the graph $G$ in Eq.~\eqref{eq:Hmaxcut} or the matrix $\bA$ 
in Eq.~\eqref{eq:Hmaxcut} are chosen by an adversary.  In other words, we want to design an algorithm 
that performs as well as possible in the worst case.
\item \emph{Algorithms from convex relaxations.} The main approach towards solving challenging optimization problems in 
computer science is based on the fact that convex optimization  can be solved efficiently, 
under minimal assumptions (roughly speaking, one has to assume that candidate solutions can be checked efficiently).  
Given a non-convex cost function (or a convex cost function over a non-convex domain),
one constructs a sequence of convex lower bounds that approximate it with increasing 
accuracy. For instance, one classical relaxation 
\cite{Goemans_Williamson_1995,Nesterov98globalquadratic} of problem \eqref{eq:Hmaxcut} is
\begin{align*}
\mbox{minimize} &  \;\;\;\; \sum_{(i,j)\in E}\<\bsigma_i,\bsigma_j\>\, ,\\
\mbox{subj. to} &  \;\;\;\; \bsigma_i\in \reals^N,\;\; \|\bsigma_i\|_2^2\, .
\end{align*}
In physics jargon, we replaced Ising spins by $N$-vector spins. This problem is non-convex 
but turns out to be equivalent to a convex one (a semidefinite program).
Remarkably any $N$-vector ground state can be `rounded' to an Ising ground state with 
universal multiplicative factor loss in energy (in the worst case).

Over the last forty years, this approach proved extremely powerful \cite{williamson2011design,barak2014sum}.
\end{enumerate}

It is also important to mention that the above points are crude simplifications. In particular, 
over the last ten years, theoretical computer scientists have been increasingly interested by the
(average case) complexity of optimizing random cost functions. One important reason
for this shift of interests
was the remarkable success of machine learning. In this context, large scale 
non-convex optimization problems (which are intractable in the worst case) are routinely
solved using gradient descent or stochastic gradient descent, see e.g. \cite{zhang2021understanding}.

Our second reference point will be statistical physics. In particular we
will revisit the classic 1986 paper by Fu and Anderson \cite{fu1986application}
on the graph partitioning problem. This requires to partition the 
vertices of a graph $G$ in two sets of equal size as to minimize the number of 
 edges that are cut by the partition. 
This can be recast as minimizing the opposite of Hamiltonian \eqref{eq:Hmaxcut}
under the constraint $\sum_i\sigma_i = 0$. In physics terms, we are minimizing a ferromagnetic 
Hamiltonian under a zero-magnetization constraint.

Let us emphasize two elements of this paper that are typical of several other papers 
in the same years:
\begin{enumerate}
\item \emph{Probabilistic model.} The authors study graph partitioning  under a random graph model, and
in particular assumes that $G$ is an Erd\H{o}s-Renyi
random graph whereby each edge is present independently with constant probability $q$. 
Under this model, they argue (non-rigorously) that the following thermodynamic limit holds
$\lim_{N\to\infty}\min_{\bsigma\in\{+1,-1\}^N}H(\bsigma)/N^{3/2} = {\sf P}_*\sqrt{q(1-q)}$ (corresponding to the max-cut size 
being $|E(G)|/2+O(N^{3/2})$), where ${\sf P}_*$ is the asymptotic ground state energy
of the Sherrington-Kirkpatrick model as given by Parisi's formula (see below). 

We notice in passing that significant 
progress was achieved on this point since  \cite{fu1986application},
and the result just mentioned was proven in \cite{dembo2017extremal}.
\item \emph{No algorithm analysis.} The paper does not propose any algorithm nor study the algorithmic question
of finding a graph partition  (a ground state $\bsigma$). Indeed, as we discuss below, the authors 
argue that statistical physics has little to offer on this crucial aspect.

As we will see, this aspect or the interaction between physics and computer science
 has evolved significantly over the last 40 years.
\end{enumerate}
%
%
%

\section{Using statistical physics to algorithmically find ground states}

The following quote from \cite{fu1986application} captures what was probably
a common viewpoint about the role of statistical physics in such interdisciplinary applications:
\begin{quote}
\emph{`In statistical mechanics we do not have complete information
about the system, nor do we demand an answer complete to the minute detail.'}

\emph{`In
optimisation problems [\dots] we are not content with a `macroscopic’ answer.'}
\end{quote}
We find here the traditional rationale for statistical physics. Namely, statistical
 physics aims at deriving
`macroscopic laws' for the behavior of matter, by abstracting away microscopic
degrees of freedom, which are `integrated over.'
The price to pay is that microscopic degrees of freedom (in particular,
the ground state for a specific realization of the Hamiltonian) seem to be lost. 

A moment of reflection shows however that probability and optimization 
are not necessarily at odds. As mentioned above, we are interested in an algorithm with the
following properties:
\begin{itemize}
\item Takes as input a realization of the energy function/Hamiltonian: $H(\; \cdot\;)$.
More precisely, the input consists of all the parameters in a specific parametrization of $H$
 (in physics jargon, the input consists of all the coupling constants).
\item It returns as output a spin configuration $\bsigma^{\salg}\in\cS^N$.
\item The output is an approximate ground state, i.e. 
$H(\bsigma^{\salg})\approx \min_{\bsigma\in \cS^N} H(\bsigma)$ with high probability. More precisely,
assuming the asymptotic ground state energy to be nonpositive:
\begin{align}
\lim_{N\to\infty}\prob\Big\{H(\bsigma^{\salg})\le  
(1-\eps_N)\min_{\bsigma\in \cS^N} H(\bsigma)\Big\} =1\,,\label{eq:ProbabilisticGuarantee}
\end{align}
where $\eps_N$ is a sequence of non-random approximation factors.
\item The running time of the algorithm is polynomial in the input size, i.e.,
in the number of parameters to specify $H$. (Since the latter is polynomial in $N$
in all the cases we are interested in, this means that the running time is polynomial in $N$.)
\end{itemize}

\begin{remark}
It is perhaps useful to discuss some aspects of the guarantee \eqref{eq:ProbabilisticGuarantee}:
\begin{enumerate}
\item The probability in Eq.~\eqref{eq:ProbabilisticGuarantee} is with respect to
the randomness in $H$ and, potentially, the additional randomness in the algorithm (when 
this is randomized). In several cases discussed below, the limit is often
approached exponentially fast, i.e. asymptotic statement could be replaced by  
$\prob(\,\cdot\,)\ge 1-2\exp(-c\, N)$ (for certain sequences $\eps_N$).
\item In a more general form of Eq.~\eqref{eq:ProbabilisticGuarantee}, we could allow 
for the multiplicative error $\eps_N$ 
as well as for an additive error, i.e. require  $H(\bsigma^{\salg})\le  
(1-\eps_N)\min_{\bsigma\in \cS^N} H(\bsigma)+\delta_N$. In all the cases discussed below,
the limit ground state energy exists (almost surely):
\begin{align}
\lim_{N\to \infty}\frac{1}{N}\min_{\bsigma\in \cS^N} H(\bsigma)=\GS\, .
\end{align}
The above definition makes sense for $\GS\le 0$ and, unless $\GS=0$, 
a multiplicative error $\eps_N$ can be traded for an additive error $\delta_N=N|\GS|\cdot \eps_N$.
\item As we will see in `mean field' models we do not expect ---in general=== to
achieve the guarantee \eqref{eq:ProbabilisticGuarantee} with $\eps_N\downarrow 0$.
In other words, there are models (random Hamiltonians) 
in which, in polynomial time,  we can only access spin configurations that are $\Theta(N)$
energy above the ground state. In these cases, $\eps$ has to be a constant independent of $N$ and
we would like to determine its minimum value (in computer science jargon, $(1-\eps)$ 
would be called the average-case 
approximation ratio).
\item Finding exact ground states requires to achieve \eqref{eq:ProbabilisticGuarantee}
with $\eps_N=0$ or $\eps_N\downarrow 0$ sufficiently fast.
\item In contrast with point 3 above, if $H$ is the Hamiltonian of a $d$-dimensional short range 
systems, under suitable conditions on the couplings distributions, we can
always achieve $\eps_N\le C(\log N)^{-1/d}$ by splitting the system in blocks of 
linear size $\ell\asymp (\log N)^{1/d}$
\end{enumerate}
\end{remark}

\subsection{Episode $\#1$: LDPC codes (1963)}

The convergence between statistical physics and optimization 
began before Fu and Anderson's paper, and even before the spin glass theory. 
To the best of my knowledge, the first step in this direction was Bob Gallager's
1963 Ph.D. thesis on low-density parity check (LDPC) codes \cite{GallagerThesis,gallager1962low}.
Gallager was not aware of the connection to statistical mechanics, which was only noticed
after the rediscovery of LDPC codes in the nineties
 \cite{kabashima1998belief,montanari2000statistical,wainwright2008graphical}.
 A review of this line of work is beyond the scope of this report 
 (see, e.g. \cite{RiU08,MeMo}).
 We will only mention a few points: 
\begin{itemize}
\item LDPC decoding can be formulated as computing marginals of a Markov Random Field
(a `graphical model') whose Markov structure is given by a sparse random graph.
\item Gallager rediscovers the  Bethe-Peierls approximation for computing these marginals.
In physics jargon, the Bethe-Peierls equations constrain the local magnetization of the system.
\item He then uses these equations as an algorithm. Namely, he iterates 
the Bethe-Peierls equations starting from an uninformative initialization (in a special, 
`message passing' parametrization). 

The output $\bsigma^{\salg}$ is obtained by choosing, for each coordinate $\sigma_i$,
the value that maximizes its marginal.
\item Gallager called this decoding method the `sum-product algorithm,' and also proposes a 
zero-temperature version of the same algorithm (which maximizes the overall likelihood),
called the `min-sum algorithm.' The sum-product algorithm was eventually rediscovered in 
the context of graphical models an artificial intelligence under the name of `belief propagation' 
(BP) \cite{pearl1988probabilistic,mceliece1998turbo}.
\end{itemize}

\subsection{Episode $\#2$: Satisfiability (2002)} 
\label{sec:FirstSAT}

In \cite{mezard2002analytic} M\'ezard, Parisi, and Zecchina studied optimization in
random $k$-satisfiability ($k$-SAT). This can be defined by a Hamiltonian of the form
\begin{align}  
H(\bsigma) = \sum_{a=1}^M\psi_a(\sigma_{i_1(a)},\dots,\sigma_{i_{k}(a)})\, ,\;\;\;\;
\bsigma\in\{+1,-1\}^N\, .\label{eq:KSAT}
\end{align}
An instance of this problem (a specific realization of the Hamiltonian) is defined
 by the $M$ $k$-uples of indices 
$i_1(a),\dots,i_{k}(a)\in [N]$, $a\in [M]$,  as well as the $M$ partial assignments 
$\bar\sigma^a_{1},\dots,\bar\sigma^{a}_{k}$. Each interaction terms in the 
Hamiltonian penalizes one assignment of the $k$ spins that appear in that term:
\begin{align}
\psi_a(\sigma_{i_1(a)},\dots,\sigma_{i_{k}(a)}) =
\begin{cases}
1 & \mbox{ if }(\sigma_{i_1(a)},\dots,\sigma_{i_{k}(a)}) = (\bar\sigma^a_{1},\dots,\bar\sigma^{a}_{k}),\\
0 &\mbox{ otherwise.}
\end{cases}
\end{align}
They study the so-called \emph{random $k$-SAT} model, whereby each of the $k$-uples
$i_1(a),\dots,i_{k}(a)$
is drawn uniformly at random and independently in $\binom{[N]}{k}$, and each of the 
partial assignments 
$(\bar\sigma^a_{1},\dots,\bar\sigma^{a}_{k})$ is drawn uniformly at random and independently in 
$\{+1,-1\}^k$. 

Inspired by the belief propagation algorithm mentioned in the previous section,
\cite{mezard2002analytic} introduced a `survey propagation'  algorithm to minimize the 
energy function\footnote{The original focus of \cite{mezard2002analytic}  was on a regime
in which the problem is, with high probability, `satisfiable' i.e. the ground state energy is zero,
but their method was generalized.} \eqref{eq:KSAT}. Their approach introduced several important 
innovations:
\begin{itemize}
\item Instead of the Bethe-Peierls approximation (which is equivalent to the replica symmetric 
cavity method), the authors adopt the one-step replica symmetry breaking (1RSB)
 cavity method. 
\item As clarified by theirs and subsequent work \cite{krzakala2007gibbs}, the effect of this change is 
that the algorithm,  does not compute marginals of the uniform measure over solutions,
as BP attempts to do. Instead, survey propagation attempts to compute marginals under a uniform measure
over clusters of solutions. 
\item Another important innovation of \cite{mezard2002analytic} is that, unlike in Gallager decoding
algorithm, coordinates $\sigma_i$ are fixed sequentially. At each step, a new coordinate is
fixed according to its estimated marginal distribution (local magnetization). `Fixing' coordinate
$i$ to $\sigma^{\salg}_i$ is equivalent to adding a term to the Hamiltonian that forces $\sigma_i$.
New local marginals are computed after each variable is fixed.
\end{itemize}

Survey propagation proved empirically successful at finding solutions of large random $3$-SAT formulas
\cite{braunstein2005survey}.
However, a precise analysis of its behavior was never carried out\footnote{A different, but related algorithm, BP-guided
decimation was analyzed (asymptotically) exactly in  \cite{montanari2007solving,ricci2009cavity}
and rigorous results were established in \cite{coja2011belief,coja2012decimation}.
A more powerful algorithm was studied in  \cite{coja2010better}. In all of these cases, for $k$ 
large, efficient algorithms fail in a large interval of the clause densities $\alpha=M/N$.}.
The initial papers suggested that the use of 1RSB cavity equations enabled this type
of algorithms to find near ground states of Hamiltonians with 1RSB-like landscapes, 
or to solve satisfiability problems with 1RSB-like space of solutions.

Subsequent rigorous work however indicates that this is unlikely to be the case,
see Section \ref{sec:Complexity}. There is now increasing evidence 
of the fact that landscapes that can be optimized are those with a full 
replica-symmetry breaking structure, as we discuss next.

\subsection{Episode $\#3$: Spin glasses and replica symmetry breaking (2019)}

It was only recently that algorithms were put forward that can \emph{provably} 
find near ground states of Hamiltonians with replica symmetry breaking
\cite{subag2021following,montanari2019optimization,el2021optimization}.
These developments took place in the context of the most canonical spin
glass model, the so called `mixed $p$-spin model.' For the sake of concreteness, we will
focus on this model, although generalizations have been developed in
\cite{el2022algorithmic,el2023local,montanari2023solving,huang2023algorithmic}.

The Hamiltonian of this model reads
\begin{align}
H(\bsigma) &= \sum_{p\ge 2}\frac{c_p}{N^{(p-1)/2}}\<\bG^{(p)},\bsigma^{\otimes p}\>\, ,
\label{eq:pspin}\\
&\bsigma\in\{+1,-1\}^N\;\;\mbox{or}\;\; \bsigma\in\S^{N-1} \sqrt{N}\, .\nonumber
\end{align}
where, the coefficients $c_p$ are independent of $N$
and, for each $p$, $\bG^{(p)}$ is a Gaussian tensor in $(\reals^N)^{\otimes p}$,
with i.i.d. entries:
\begin{align}
\bG^{(p)} =(G^{(p)}_{i_1\dots i_p})_{i_1\dots i_p\le N}\sim_{iid}\normal(0,1)\, ,
\label{eq:GaussianCouplings}
\end{align}
It is customary to encode the coefficients  $c_p$ in the generating 
function $\xi(q) :=\sum_{p\ge 2}c_p^2 q^p$, and to assume these coefficients to decay fast enough
that $\xi(q)<\infty$ for some $q>1$. 
In fact there is little loss of generality in assuming $c_p=0$ for all $p>p_{\max}$
(for some constant $p_{\max}$). 

The energy function \eqref{eq:pspin} is known to be NP-hard to approximate (in the worst case),
except in the special case of spherical constraint (i.e. $\bsigma\in\S^{N-1} \sqrt{N}$)
and quadratic energies (i.e. $c_p\neq 0$ only for $p=2$). The quadratic Ising case is
the celebrated Sherrington-Kirkpatrick model \cite{sherrington1975solvable}, and
is hard to approximate (in the worst case) within a logarithmic 
factor\footnote{This means that there exists a constant $c_0>0$  such that, unless P$=$NP, 
 no polynomial algorithm is guaranteed to achieve $H(\bsigma^{\salg})\le 
\min_{\bsigma}H(\bsigma)/(\log N)^{c_0}$ for every matrix $\bG^{(2)}$.} \cite{arora2005non}.

\begin{figure}[t]
  \phantom{A}\hspace{1cm}
  \begin{tikzpicture}[thick, level distance=8mm, inner sep = 0.2mm]
    \draw[very thick, ->] (-4,0) -- (-4,-4);
    \node at (-3.75,-2) [rotate = 90] {$t=q$};
    \node at (1,-1.7) {$\gamma$};
    \node at (0.3,-4.2) {$\alpha_1$};
    \node at (1.35,-4.2) {$\alpha_2$};
  \tikzstyle{level 1}=[sibling distance=30mm]
  \tikzstyle{level 2}=[sibling distance=15mm]
  \tikzstyle{level 3}=[sibling distance=7.5mm]
  \tikzstyle{level 4}=[sibling distance=3.75mm]
  \tikzstyle{level 5}=[sibling distance=1.875mm]
   \node at (0,0) [circle,draw=black!100,fill=black!100] {}
    child foreach \x in {0,1}
      {child foreach \y in {0,1}
        {child foreach \z in {0,1}
          {child foreach \w in {0,1}
            {child foreach \u in {0,1}}} }};
  \tikzstyle{level 1}=[sibling distance=20mm, level distance=22mm]
  \tikzstyle{level 2}=[sibling distance=7.5mm, level distance=6mm]
  \tikzstyle{level 3}=[sibling distance=3.75mm, level distance=6mm]
  \tikzstyle{level 4}=[sibling distance=1.875mm, level distance=6mm]
  \node at (7,0) [circle,draw=black!100,fill=black!100] {}
     child foreach \x in {0,1,2}
      {child foreach \y in {0,1}
        {child foreach \z in {0,1}
          {child foreach \w in {0,1}} }};
      \draw [dashed] (4,0) -- (10,0);
      \draw [dashed] (4,-2.2) -- (10,-2.2);
      \draw [dashed] (4,-4) -- (10,-4);
      \node at (9.7,-0.3) {$q=0$};
      \node at (9.7,-2.5) {$q_0$};
      \node at (9.7,-4.3) {$q_*$};
    \end{tikzpicture}
    \caption{Cartoon of the tree of ancestor states: two examples. Tree levels are indexed by the overlap value $q$,
      which corresponds to the time index in the algorithm evolution $t$, and to the norm of magnetization vectors
      $\|\bm^{\gamma}\|_2^2/N=q$ (if $\gamma$ is at level $q$). On the right: a tree with 
      large overlap gap between $0$ and $q_0$. (From \cite{alaoui2020algorithmic}.)}
    \label{fig:TreeCartoon}
  \end{figure}
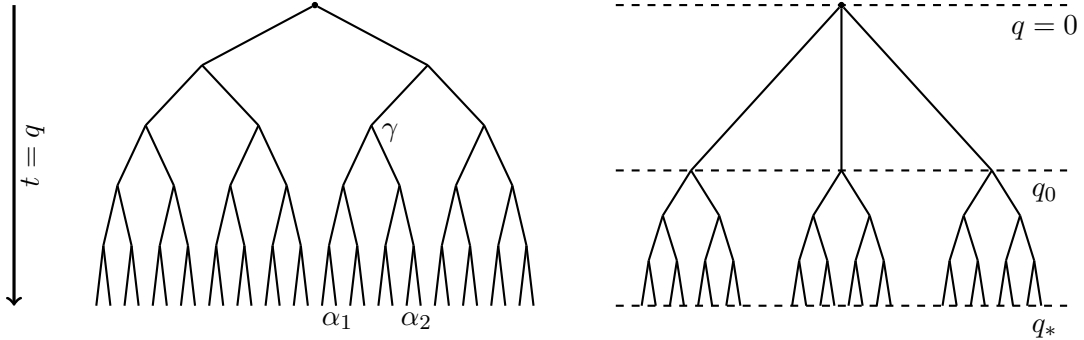

I will not describe in any detail the algorithms of 
\cite{subag2021following,montanari2019optimization,el2021optimization},
but rather try to convey the intuition of how the success of these algorithms is 
connected to the replica-symmetry breaking  structure of the energy landscape.
This landscape structure was clarified in a sequence of classicaal papers in 
spin glass theory \cite{mezard1984nature,mezard1985microstructure}
and ---to a remarkable extent--- made rigorous in recent years 
\cite{panchenko2013parisi,panchenko2013sherrington,chen2018generalized,chen2019generalized}. 

The first fact discovered in these papers is that mean field spin glass
 models have a large (diverging with $N$) 
number of near ground states\footnote{I will be a bit loose about the definition of `near ground
state,' but the reader can take it to mean configurations $\bsigma$ with 
$H(\bsigma)\le \min_{\bsigma'}H(\bsigma')+N\eps$ with $\eps$ a small constant or
$\eps\downarrow 0$ slowly (in particular $N\eps\uparrow \infty$).} $\bsigma^{\alpha}\in\Sigma_N$
(here $\Sigma_N=\{+1,-1\}^N$ or $\S^{N-1} \sqrt{N}$), 
which are at large distances from each other,
namely $\|\bsigma^{\alpha}-\bsigma^{\beta}\|^2\ge cN$, for some constant $c>0$.
If we denote by\footnote{Here $\conv(A)$ denotes the convex envelope of set $A$.} $\bm^{\circ}\in \conv(\Sigma_N)$ the barycenter of these near ground-states,
then $\<\bsigma^{\alpha}-\bm^{\circ},\bsigma^{\beta}-\bm^{\circ}\>  = o(N)$. 
The second crucial discovery of \cite{mezard1984nature,mezard1985microstructure} is that these 
states are organized according to a (balanced) tree, rooted at $\bm^{\circ}$ with leaves at the $\bsigma^{\alpha}$.
Each internal node $\gamma$ of the tree can be labeled with the barycenter of the 
configurations (leaves) that appear among its descendants). A consequence of this construction
is that, if $\gamma_1\preceq \gamma_2 \preceq \gamma_3$ (where $\preceq$ means `is an ancestor of'),
then $\<\bm^{\gamma_2}-\bm^{\gamma_1},\bm^{\gamma_3}-\bm^{\gamma_2}\>= o(N)$.
Finally, the tree has the property that all nodes $\gamma$ at level $\ell$
(graph distance $\ell$ from the root) 
have approximately the same norm $\|\bm^{\gamma}\|^2 = Nq_{\ell}+o(N)$.
The level of the leaves is indicated by $L$.

Now, two situations are possible, as portrayed in the cartoon of Figure
\ref{fig:TreeCartoon}. Either the tree branches continuously, namely $L\to\infty$
and $q(\ell+1)-q(\ell) \to 0$ for all $\ell\le L$ as $N\to\infty$, or there is a gap,
namely an $\ell_0$ (possibly diverging) such that $q(\ell_0+1)-q(\ell_0)$ remains bounded 
away from $0$ as $N\to\infty$.
We say that in the first case there is  `no overlap gap' and in the second 
there is an `overlap gap.' These two scenarios can be distinguished by the absence or presence of a gap
in the support of the asymptotic overlap distribution.

Notice that in the no-overlap gap case there must necessarily be full RSB,
while in the case of finitely many steps of RSB there must be an overlap gap. On the other hand, 
we can have full RSB and overlap gap. Indeed this expected to be the case (for instance) for pure $p$-spin 
($p\ge 3$) Ising models (i.e., the case in which $c_k\neq 0$ only for $k=p$).

\begin{figure}
\begin{center}
\includegraphics[width=7cm,angle=0]{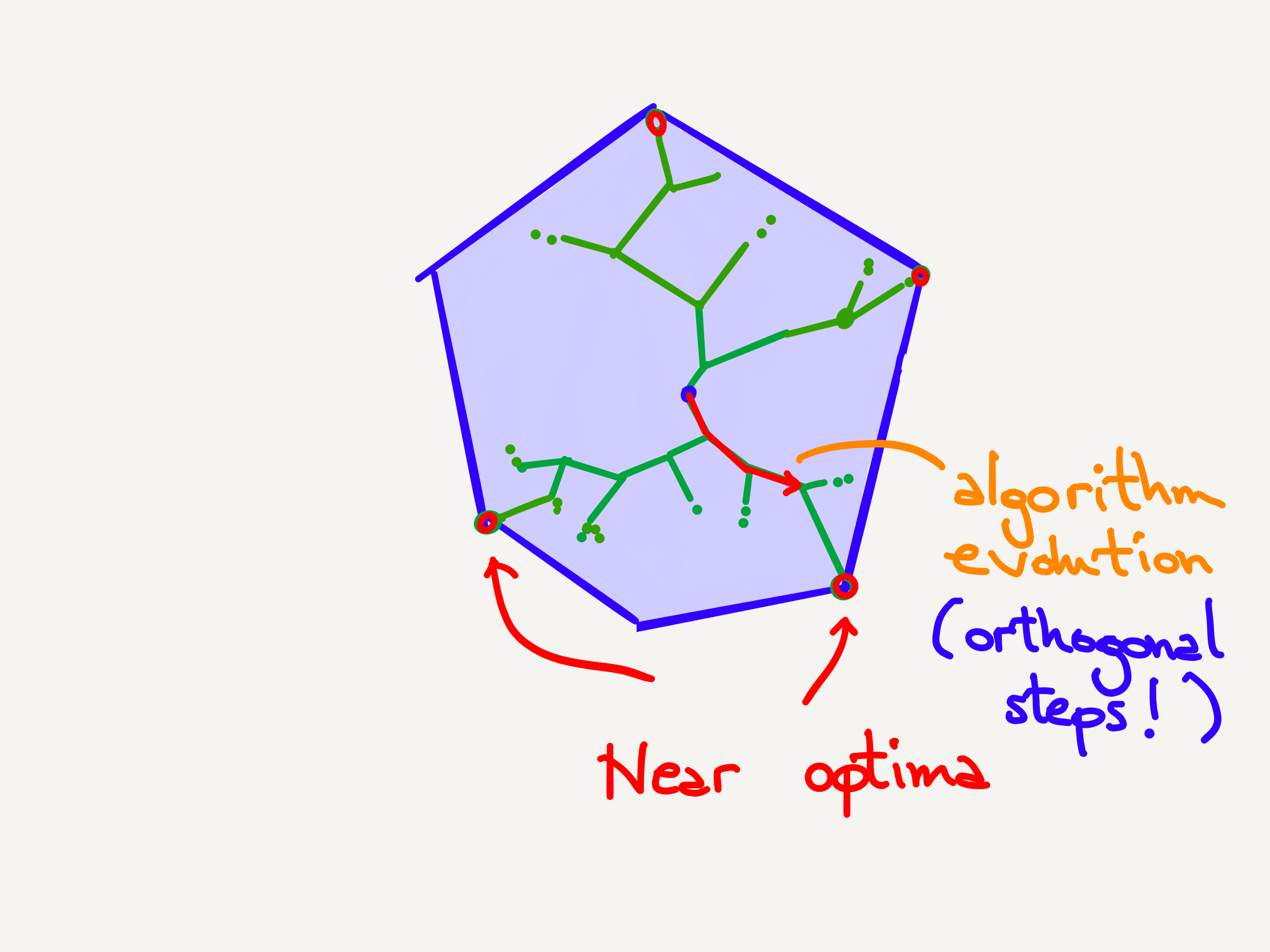}
\end{center}
\caption{A sketch of the behavior of the algorithms of 
\cite{montanari2019optimization,el2021optimization}.
Starting from the center of the hypercube., the algorithm follows a branch of the tree
that organizes near-ground states.}\label{fig:SketchOpt}
\end{figure}

In the case of no overlap gap the distance between the barycenters $\bm^{\gamma_1}$, $\bm^{\gamma_2}$
associated to neighboring nodes is `small'. Indeed, if $\gamma_2$ is at level $\ell+1$
and $\gamma_1$ at level $\ell$, the $\|\bm^{\gamma_2}-\bm^{\gamma_1}\|^2=N(q(\ell+1)-q(\ell))+o(N)=o(N)$.
Hence, we can hope to move from one level to the next one by 
optimizing locally a suitable potential function. If this is possible, then we can perhaps
reach a near-ground state (a leaf of the tree) by descending the tree of ancestor states, 
and making at each step a decision based on the local landscape,
see Figure \ref{fig:SketchOpt} for a cartoon.

Indeed, this is the strategy followed by the algorithm of
\cite{subag2021following,montanari2019optimization,el2021optimization}.
Consider for the sake of simplicity the case in which the barycenter is $\bm^{\circ}=\bfzero$
(this is the case of model \eqref{eq:pspin} with no magnetic field, i.e. no term linear in $\bsigma$).
These algorithms construct a sequence of estimates $\bm^{t}$,
with $t\in\{0,\delta,2\delta,\dots,1\}$, such that $\|\bm^{t}\|^2=Nt$.
at each step, the increment $\bm^{t+\delta}-\bm^t$ is chosen to be orthogonal to $\bm^t$,
and indeed can be chosen to be orthogonal to all previous iterates.
The stepsize $\delta$ is let vanish after $N\to\infty$.
(See Figure \ref{fig:SketchOpt} for a sketch.)

Of course the way in which the increment $\bm^{t+\delta}-\bm^t$ is chosen is important,
and differs between the spherical \cite{subag2021following}
and the Ising cases \cite{montanari2019optimization,el2021optimization}:
\begin{itemize}
\item In the spherical case, \cite{subag2021following} extends $H(\bsigma)$ in the natural way to
a function on $\conv(\Sigma_N)  = B^N(\sqrt{N})$ (the Euclidean ball of radius $\sqrt{N}$).
At  step $t$, $\bm^{t+\delta}-\bm^t$ is chosen to minimize the second order term 
of the Taylor expansion of $H(\,\cdot\,)$ around $\bm^t$, in the orthogonal complement of $\bm^{t}$.
\item In the Ising case, a naive application of this method fails, if nothing else
because it does not account for the hypercube structure and $\bm^t$ exits
$\conv(\Sigma_N) = \{\bm\in\reals^N:\,\|\bm\|_{\infty}\le 1\}$.
The papers \cite{montanari2019optimization,el2021optimization} choose the next iteration using
a general `approximate message passing' (AMP) procedure \cite{bayati2011dynamics,javanmard2013state}. 
This type of procedure
can be analyzed precisely in the $N\to\infty$ limit, and 
\cite{montanari2019optimization,el2021optimization} choose the precise form of the update
as to optimizing the resulting energy.

Applied to the spherical model, this approach recovers the same energy achieved by the 
Hessian algorithm of \cite{subag2021following}.
\end{itemize}

In the case in which the overlap distribution exhibits no overlap
gap, then these algorithms can be shown to achieve a near ground state.
Namely, Eq.~\eqref{eq:ProbabilisticGuarantee} holds for some $\eps_N\downarrow 0$.
This is expected to be the case for the Sherrington-Kirkpatrick model, since in this case it is 
expected (but not proven) that the overlap distribution has no overlap gap.
\begin{remark}
Let is emphasize once more that the fact that  Eq.~\eqref{eq:ProbabilisticGuarantee} holds for some $\eps_N\downarrow  0$
 does not mean that exact ground states can be found in polynomial time. 
 For instance, it is in principle possible that the complexity of finding an $\eps$-ground state
 (i.e. satisfying Eq.~\eqref{eq:ProbabilisticGuarantee})
 is of order $\exp(N^{a}c(N^b\eps)+O(\log N))$,
  with $c(0)>0$ a constant and $c(x)\asymp x^{-a/b}$ as $x\to\infty$.
  \end{remark}

Even more interestingly, \cite{el2021optimization} characterizes 
the optimal energy achieved by this class of algorithms. In order to
state this result, it is useful to recall Parisi's formula for the ground state energy.
Given a function $\gamma:[0,1]\to \reals_{\ge 0}$, consider the following partial differential equation,
also known as the Parisi PDE:
\begin{align}\label{eq:PDEFirst}
\begin{split}
\partial_t \Phi_{\gamma}(t,x)+\frac{1}{2}\xi''(t) \Big(\partial_x^2\Phi_{\gamma}(t,x)+\gamma(t) (\partial_x\Phi_{\gamma}(t,x))^2\Big) = 0, ~~~ (t,x)\in [0,1)\times \reals \, ,
\end{split}
\end{align}
with boundary condition $\Phi_{\gamma}(1,x) = |x|$ (in the Ising case)
or $\Phi_{\gamma}(1,x) = x^2/2$ (in the spherical case). It is known that a weak solution 
to the above PDE exists and is unique if $\gamma\in\cuL$ 
\cite{jagannath2016dynamic,el2021optimization}, where we define the following spaces of functions
$\cuU\subseteq \cuL$:
\begin{align*}
 \cuU&:=\Big\{\gamma:[0,1)\to\reals_{\ge 0}\;\;\;\;\mbox{non-decreasing} \Big\}\, ,\\
\cuL &:= \Big\{\gamma:[0,1)\to \reals_{\ge 0}: \;\; \|\xi''\gamma\|_{\sTV[0,t]}<\infty~ \forall t\in [0,1), \int_0^1\!\xi''\gamma(t)\,\de t < \infty\Big\} \, .
\end{align*}
Here $\xi''\gamma(t):=\xi''(t)\gamma(t)$, and $ \|\xi''\gamma\|_{\sTV[0,t]}$ is the total 
variation of this function over the interval $[0,t]$. In particular, the condition
 $\|\xi''\gamma\|_{\sTV[0,t]}<\infty$ is satisfied if,
 in  the interval $[0,t]$,  $\xi''\gamma$ has a finite number 
 of jump discontinuities and is Lipschitz continuous outside these 
 singularities.

We use
the solution of this PDE to define the following variational functional
\begin{align}
  \Par(\gamma) &:= -\Phi_{\gamma}(0,0)+\frac{1}{2}\int_{0}^1 t\xi''(t)\gamma(t)\,
   \de t\, ,\label{eq:ParisiFunctional}
\end{align}

Parisi's formula expresses the  asymptotic ground state energy in terms of
a variational principle over $\cuU$:
\begin{align} 
\lim_{N\to\infty}\frac{1}{N}\min_{\bsigma}H(\bsigma) &= \;\GS\phantom{E}:= 
\sup_{\gamma\in\cuU} \sP(\gamma)\, .
\end{align}
The proof of this result in
\cite{talagrand2006parisi,panchenko2013sherrington,auffinger2017parisi} was a major
achievement in probability theory. Further, $\gamma\mapsto \Par(\gamma)$ is known to be strictly concave
\cite{auffinger2015parisi,jagannath2016dynamic} and therefore the
supremum is achieved at a unique $\gamma_{\GS}$ which is interpreted to 
(a rescaling of) the asymptotic overlap cumulative distribution function.
In particular, the no-overlap gap condition corresponds to $\gamma_{\GS}$ being strictly increasing.

Remarkably, the optimum energy achieved by the algorithms outlined above can be expressed in 
terms of a modified variational principle.
  \begin{theorem}[\cite{el2021optimization}]\label{thm:ALG}
 Assume that the supremum $\sup_{\gamma\in\cuL} \sP(\gamma)$ is achieved at some $\gamma_{\salg}
 \in \cuL$. Then there exists an algorithm with complexity $O(N^{p_{\max}}\log N)$
 which outputs $\bsigma^{\salg}\in \Sigma_N$ (either $\Sigma_N= \{+1,-1\}^N$
 or $\Sigma_N= \S^{N-1}\sqrt{N}$) such that
\begin{align*}
\lim_{N\to\infty}\frac{1}{N}H(\bsigma^{\salg}) &=\ALG:= \sup_{\gamma\in\cuL} \sP(\gamma)\, .
\end{align*}
In particular, $\ALG=\GS$ if and only if $\gamma_{\GS}$ is strictly increasing.
\end{theorem}
We thus recovered the claim that we can optimize FRSB energy landscapes 
with no-overlap gap, and additionally characterized the energy we can achieve by a class of polynomial
time algorithms. Below, we will discuss evidence that $\ALG$ is
indeed a fundamental computational barrier.
%
%
\section{From statistical physics to average-case complexity}
\label{sec:Complexity}

Let us return to Fu and Anderson \cite{fu1986application}. Here is another 
interesting quote:
\begin{quote}
\emph{`\dots  phase transitions will affect the actual implementation and performance of such
algorithms.'}

\vspace{0.5cm}

\emph{`\dots and the accompanying knowledge of the structure of
solution space may also play an important role in complexity theory'}
\end{quote}
The authors conjecture that phase transitions and the geometry of the space of solutions
(or, we could add, near optima) will play a role in determining the complexity of the 
optimization problem. However, it is fair to say that they do not provide any
intuition or heuristics about how this connection might arise or why.

In physics, one is tempted to think of algorithms as generalizations of 
physical dynamics, and hence the above intuition might have been motivated by the impact of 
phase transitions on such dynamics.
However --in general-- the space of algorithms
is much larger than the space of natural physical dynamics
(admittedly, we make no attempt at formalizing the last notion). For instance, solving linear systems over a finite field can be done efficiently
via Gaussian elimination, but is very hard for natural physical dynamics \cite[Chapter 18]{MeMo}.

It is all the more surprising that the prediction quoted above turned out to be correct.
The most interesting part of the story is of course in the how and why.

\subsection{Episode $\#1$: Satisfiability (1999-2002)}

\begin{figure}
\hspace{3cm}\includegraphics[width=3.8cm,angle=-90]{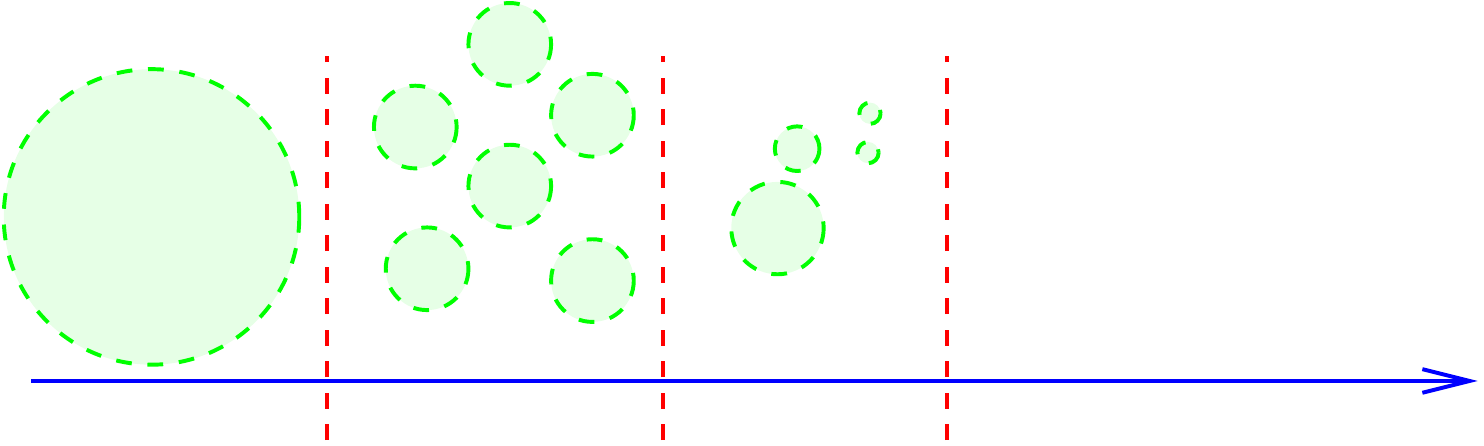}
\put(-275,-103){{\small $\alpha_{\rm d}(k)$}}
\put(-195,-103){{\small $\alpha_{\rm c}(k)$}}
\put(-125,-103){{\small $\alpha_{\rm s}(k)$}}
\end{figure}

The first important steps towards clarifying the connection between phase transitions, 
geometry of the solution space, and computational complexity 
were achieved again in the context of $k$-satisfiability ($k$-SAT), in
\cite{monasson1999determining,mezard2002analytic}.
Both papers studied random $k$-SAT, in the limit $M,N\to\infty$,
with $M/N\to\alpha$, cf. Section \ref{sec:FirstSAT}. 
Random instances undergo a phase transition as $\alpha$ crosses a critical value $\alpha_{s}(k)$,
see\footnote{The statistical physics calculation of this threshold was carried out in
\cite{mezard2002analytic}, and the existence of a threshold in this sense follows from
the existence of thermodynamic limit proven in \cite{franz2003replica_2}, see also
\cite{franz2003replica,panchenko2004bounds}.}
 \cite{mezard2002analytic,franz2003replica_2}:
\begin{align}
\alpha<\alpha_{\rm s}(k)\;\; \Rightarrow \;\; \lim_{N\to\infty}\frac{1}{N}\min_{\bsigma\in\{+1,-1\}^N}
H(\bsigma) = 0\, ,\\
\alpha>\alpha_{\rm s}(k)\;\; \Rightarrow \;\; \lim_{N\to\infty}\frac{1}{N}\min_{\bsigma\in\{+1,-1\}^N}
H(\bsigma) > 0\, .
\end{align}
In fact the phase transition is believed to be stronger. Namely, for $\alpha<\alpha_{\rm s}(k)$
not only the ground state energy is sub-extensive, but it is exactly zero with high probability:
 $\prob(\min_{\bsigma\in\{+1,-1\}^N}H(\bsigma) = 0)\to 1$. One says that the instance is 
`satisfiable' in this case. This satisfiability conjecture in known to hold for $k=2$
\cite{chvatal1992mick} (by a mapping of this problem onto a random graph problem), and was recently proved
for all $k\ge k_*$ for a suitably large $k_*$ \cite{ding2022proof}. The latter result was a 
real breakthrough in probability theory,
largely inspired by ideas from spin glass theory.

Coming back to the computational complexity question, the papers
\cite{kirkpatrick1994critical,monasson1999determining} focused on the problem of finding 
`satisfying assignments', i.e. zero energy configurations when such configurations exist.
They brought attention to two intriguing phenomena:
\begin{enumerate}
\item They empirically showed that, for many satisfiability solvers, the hardest random instances
are the ones generated with $\alpha$ close to $\alpha_{\rm s}(k)$. 
\item They defined a suitable order parameter 
for the phase transition at $\alpha_{\rm s}(k)$: the fraction of spins that are `frozen,' i.e. take the
same value in all the ground states.
They argued that for $k=2$ this order parameter is continuous at the phase transition, 
while for $k\ge 3$ it is discontinuous.

 They pointed at an  intriguing correspondence with the distinction\footnote{This distinction applies
 to the satisfiability version of the problem, whereby we request to determine whether 
 a satisfying assignment (a zero-energy configuration) exists and to find one, or to prove that
 it does not exist. The optimization version (minimizing the energy) is NP hard also for $k=2$.}
in complexity between satisfiability for $k=2$ (which is polynomial) and $k\ge 3$ (NP-complete). 
\end{enumerate}

These findings acquired further depth in \cite{mezard2002analytic},
which --among other results-- used the 1RSB cavity method to first compute the exact 
threshold $\alpha_{\rm s}(k)$.
More importantly for our discussion, \cite{mezard2002analytic} 
characterized the so-called `shattering' phase transition\footnote{The name `shattering'
was introduced in \cite{achlioptas2008algorithmic} while earlier papers used the term `clustering.'}
(see also \cite{biroli2000variational} for earlier work and \cite{krzakala2007gibbs,achlioptas2008algorithmic,
achlioptas2011solution,sly2016number} for subsequent work),
that takes place at a point $\alpha_{\rm d}(k)<\alpha_{\rm s}(k)$. 

The picture suggested by the physics calculations was the following.  
For $\alpha>\alpha_{\rm d}(k)$, the set of solutions (zero-energy ground states)
shatters into exponentially many clusters, with any two clusters separated by a 
Hamming distance larger than $Nc$, for some $c>0$. In contrast, for 
 $\alpha<\alpha_{\rm d}(k)$ most solutions are `well connected\footnote{In particular
most pairs of solutions $\bsigma$, $\bsigma'$ can be connected by a path of solutions 
$\bsigma^{(0)}=\bsigma,\bsigma^{(1)},\dots,\bsigma^{(L)}=\bsigma'$,
with `small steps' $\|\bsigma^{(\ell)}-\bsigma^{(\ell+1)}\|^2=o(N)$.}.'
From the point of view of spin glass theory, $\alpha_{\rm d}(k)$ marks a dynamical
1RSB phase transition.
Several elements of this picture have been proved since, either in the context of
random $k$-SAT or for closely related models.

The authors of  \cite{mezard2002analytic} suggested that this
dramatic change in the geometry of the solution space impacted algorithms,
making random instances hard to solve. At the same time they expected survey
propagation to bypass this obstruction because it could account for 
long range correlations arising in a 1RSB model. 

As we will see next, the 1RSB structure 
(or more generally, overlap gap) can indeed be  used to prove hardness, but 
the resulting threshold does not coincide precisely with the dynamical phase transition\footnote{On the other
hand, there is rigorous evidence that the dynamical phase transition marks the onset of hardness for sampling,
see Section \ref{sec:Conclusion}.}. Also, we do not expect that this fundamental obstruction 
can be eluded by accounting for RSB in the optimization algorithm.

\subsection{Episode $\#2$: Hardness from overlap gap (2013)}

The next breakthrough was achieved by Gamarnik and Sudan \cite{gamarnik2014limits}
who were able to use geometric properties of the set of solutions to prove that 
a broad family of algorithms fails to find near optimizers. The focus of the original paper
was on maximum-size independent sets but, since then, their approach was extended
with several technical innovations, and applied to a broad variety of problems.
A recent overview of this line of work is given in \cite{gamarnik2021overlap}.

I will try to describe the basic intuition of \cite{gamarnik2014limits}
in very general terms, skipping mathematical formalizations.
As before we have a random Hamiltonian $H$ and, to be definite, we assume that 
the configuration space is the discrete hypercube $\{+1,-1\}^N\ni \bsigma$.
We will think that the minimum/maximum of $H$ are of order $N$, 
with the minimum being negative, as is customary in physics.
The basic elements for such hardness results are:
\begin{itemize}
\item A definition of what is a `near-ground state.' One possible choice is to 
fix some (small) constant $\eps$ and define
\begin{align}
\cuE(\eps) := \Big\{\bsigma\in \{+1,-1\}^N:\; H(\bsigma) \le  (1-\eps)\min_{\bsigma'\in\{+1,-1\}^N}
H(\bsigma')\Big\}\, .\label{eq:sublevel}
\end{align} 
(If $\min_{\bsigma'\in\{+1,-1\}^N}
H(\bsigma') = N\GS+o_P(N)$  for some constant $\GS<0$, then the above is roughly 
equivalent to $H(\bsigma) \le  \min_{\bsigma'\in\{+1,-1\}^N}
H(\bsigma')+N|\GS|\eps$.)

Let us emphasize that $\cuE(\eps)$ is a non-empty set and depends on the Hamiltonian $H$.
Since $H$ is random, $\cuE(\eps)$ will be itself random.
\item  As definition of a suitable class of algorithms.
Any specific problem is defined by a class of Hamiltonians, which we will imagine  
parametrized by a vector of parameters $\bg\in\reals^M$. For instance, in
the case of the spin glass model \eqref{eq:pspin}, $\bg = (\bG^{(p)}/n^{(p-1)/2})_{2\le p\le p_{\max}}$
is the collection of all couplings, and $M = N^2+N^3+\cdots+N^{p_{\max}}$.

We then think of an algorithm as a function that takes as input the 
parameters (the couplings) and returns as output a spin configuration:
\begin{align}
\bF:\reals^M\to \{+1,-1\}^N\, .
\end{align}
The crucial assumption is that the output of the algorithm depends continuously on its input,
i.e. $\bF$ is a continuous function, in a suitably quantitative sense.

With an abuse of notation, we will think $\bF$ as a function of the Hamiltonian itself
$H\mapsto \bF(H)$.
\end{itemize}

Next we draw two independent random Hamiltonians $H_0$ and $H_1$ from the
same distribution as our original Hamiltonian $H$, and
we construct a continuous path of Hamiltonians  $(H_t)_{0\le t\le 1}$ which interpolates between 
$H_0$ and $H_1$. The path will be such that, for each $t\in [0,1]$, the Hamiltonian $H_t$ has the same 
distribution\footnote{These requirements can be somewhat relaxed, but we are not trying to
present the most general version here.} as $H_0$ and $H_1$.

\begin{figure}
\begin{center}
\includegraphics[width=5cm,angle=0]{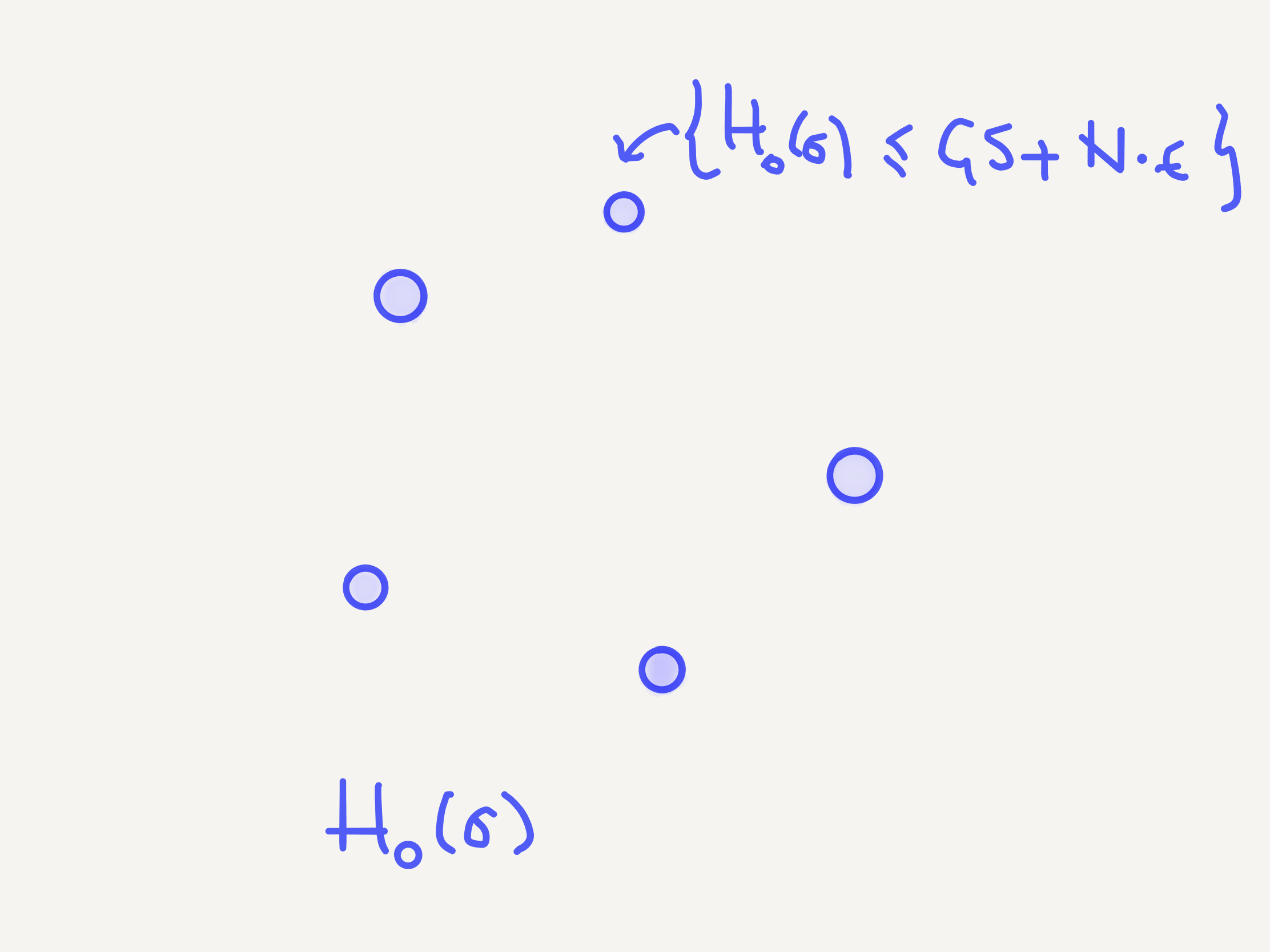}
\includegraphics[width=5cm,angle=0]{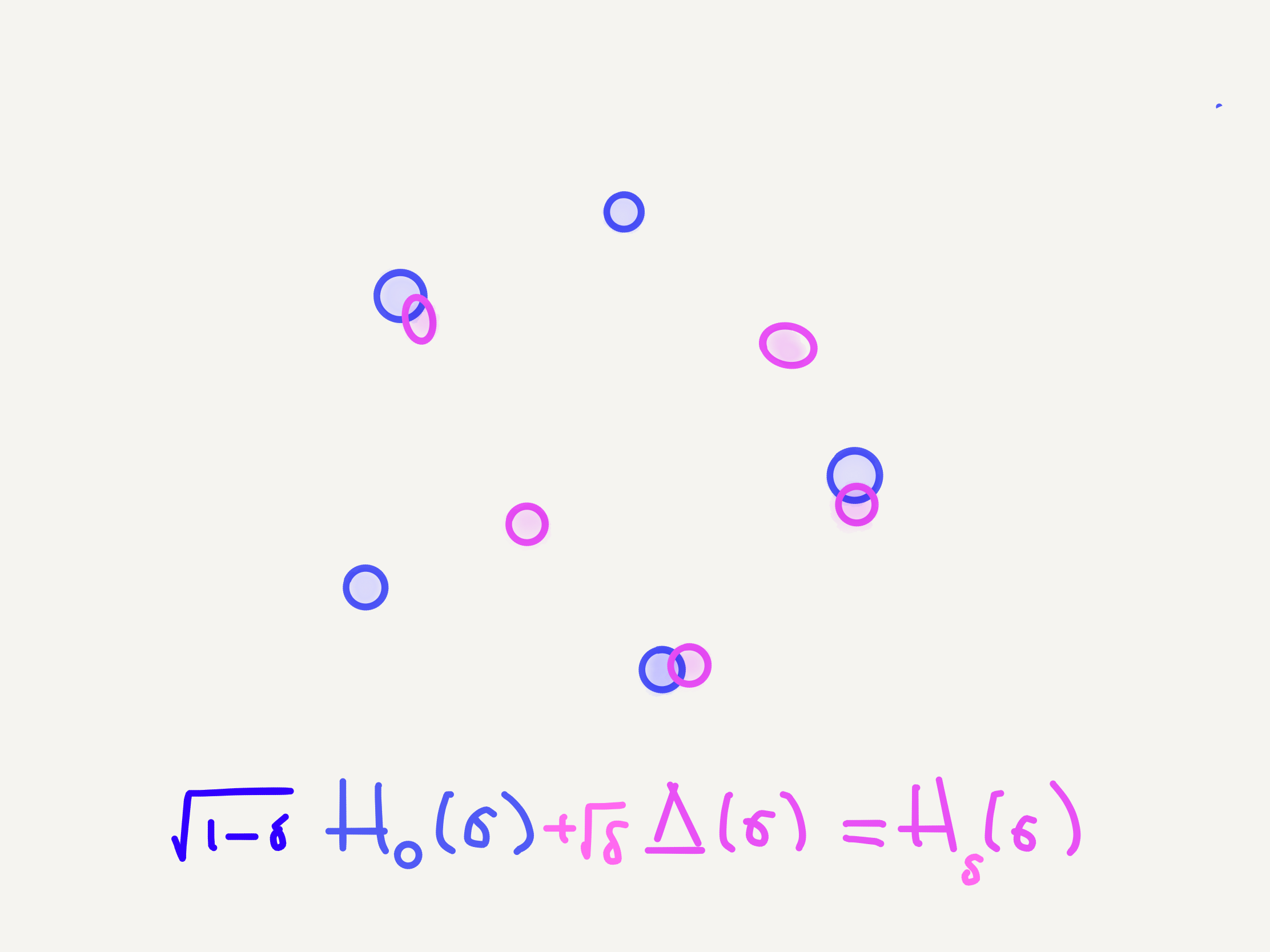}
\includegraphics[width=5cm,angle=0]{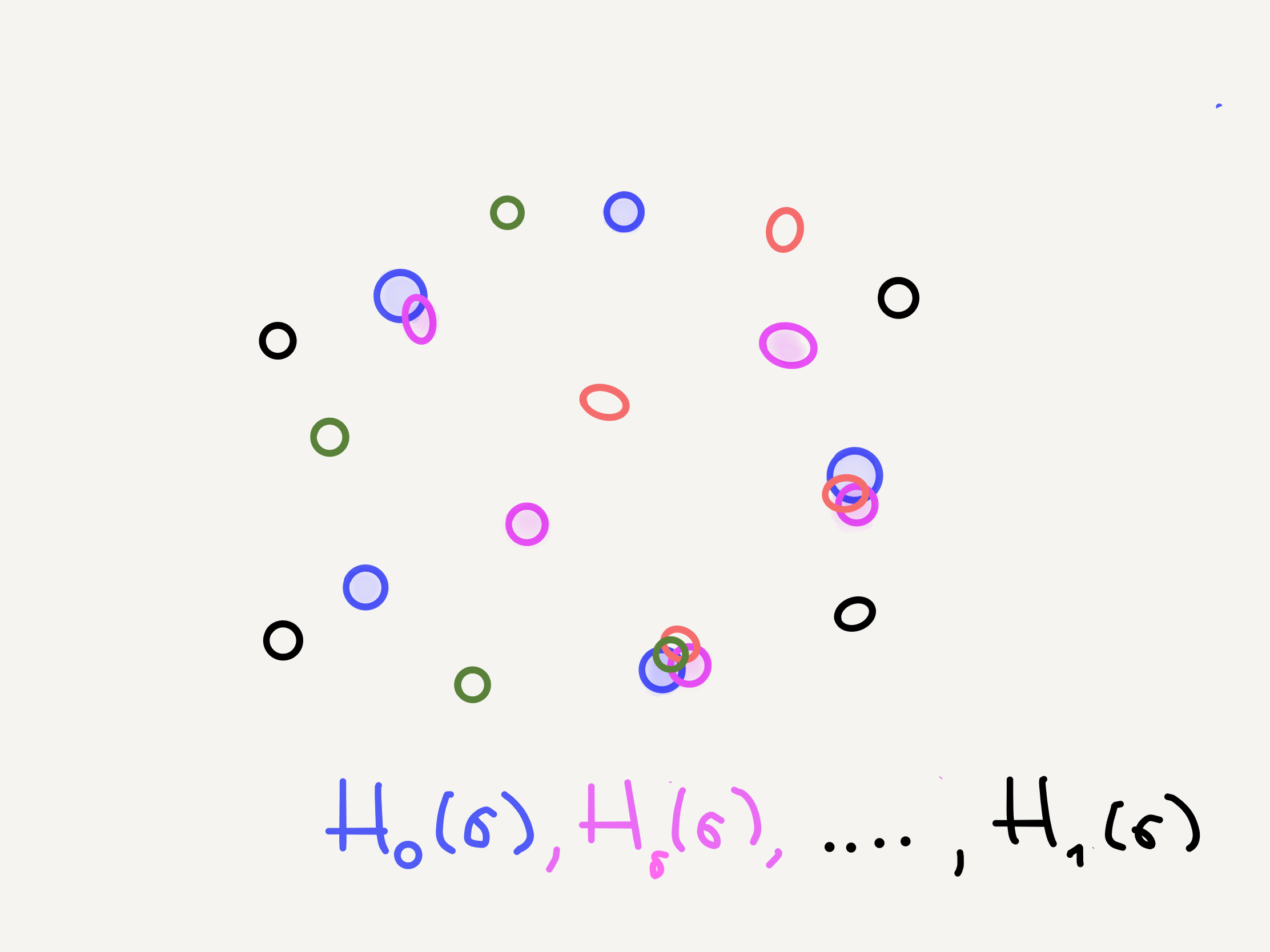}
\end{center}
\caption{Sketch of the evolution of the set of near-ground states 
along the interpolation path between two i.i.d. Hamiltonians.}
\end{figure}
For instance, in the case of the spin glass model \eqref{eq:pspin}, with Gaussian 
couplings, cf. Eq.~\eqref{eq:GaussianCouplings}, $H_0$ and $H_1$  could be defined using
two i.i.d. realizations of the couplings $\bG^{(p,0)}$, $\bG^{(p,1)}$, and $H_t$
can be defined via
\begin{align}
H_t(\bsigma) = \sqrt{1-t}\, H_0(\bsigma)+\sqrt{t}\, H_1(\bsigma)\,.
\end{align}
We will denote by $\cuE(\eps;t)$ the set of near ground states for $H_t$,
which is defined as per Eq.~\eqref{eq:sublevel}, with $H$ replaced by $H_t$.

How does $\cuE(\eps;t)$ look like in a model with overlap gap (e.g. in a 1RSB scenario)?
For any fixed $t$, the law of $\cuE(\eps;t)$ is the same as the one of  $\cuE(\eps)$.
In particular, with high probability, it will consist of a large number of connected components:
\begin{align}
\cuE(\eps;t)=\bigcup_{i=1}^{J(t)}\cuC_i(\eps;t)\, ,\label{eq:Decomposition}
\end{align}
 with $\dist(\cuC_i(\eps;t),\cuC_j(\eps;t))\ge N\delta_0$
for some strictly positive constant (here $\dist$ is Hamming distance or, equivalently,
square Euclidean distance). We will refer to these well-separated connected components
as `states,' since they correspond to ancestor states in the tree decomposition of 
the Gibbs measure.

How does the set $\cuE(\eps;t)$ evolve with $t$? In our spin glass example, cf. Eqs.~\eqref{eq:pspin},
\eqref{eq:GaussianCouplings}, moving from $t_0$ to $t_0+\delta$ is equivalent to
changing the Hamiltonian from $H_{t_0}$ to $\sqrt{1-\delta}H_{t_0}+\sqrt{\delta}\Delta$
with $\Delta$ an independent copy of $H$. This perturbation can have multiple effects on
states. It can make a state $\cuC_i(t_0;\eps)$ `disappear' (because it raises its energy above the threshold
defined by $\eps$), it can make a state (which did not exist at $t_0$) `appear'
as $\cuC_i(t_0+\delta;\eps)$, or it can `modify continuously' $\cuC_i(t_0;\eps)$
into $\cuC_i(t_0+\delta;\eps)$. Finally a states can also `split' into multiple ones
or `merge'. We call a state a `survivor' if it can be traces continuously between $t=0$
and $t=1$ (with some choices at splitting times.)

Now assume algorithm $\bF$ is successful i.e., with high probability, on
input $H$, it returns $\bsigma=\bF(H)\in \cuE(\eps)$. Let $\bsigma_t = \bF(H_t)$ 
be its output on input $H_t$. Then $t\mapsto \bsigma_t$ is a path in the space of configurations
which we expect to have the following properties (proving that this is the case requires
careful technical work): $(1)$~For each $t\in [0,1]$ we have $\bsigma_t\in \cuE(\eps;t)$;
$(2)$~The path $t\mapsto \bsigma_t$ is continuous (because $t\mapsto H_t$ and $H\mapsto \bF(H)$
are). Put together, these properties imply that there must be at least one survivor state.

So far, the argument was very general: we did not invoke any specific property of
the random Hamiltonian $H$, apart from the overlap gap property, which implies the 
decomposition \eqref{eq:Decomposition}. At this point, one makes a model-specific 
computation and proves that, for the model under consideration and the specific approximation
ratio $\eps$, there are no survivor states. 
This is agreement with our intuition about 1RSB landscape: each perturbation shifts the energies of
different states in a roughly independent fashion. As a consequence,
states will keep appearing and disappearing from the decomposition 
\eqref{eq:Decomposition} as their energy shifts below or above the approximation level $\eps$.

\subsection{Episode $\#3$: Sharp results for mixed $p$-spin models (2022)}

The proof technique outlined above does not always work and, when it
does, does not always produce sharp hardness results. Several refinements of
the original idea have been developed in recent years. Arguably, the most 
sophisticated of such techniques is the one based on the 
`branching overlap gap property' introduced by Huang and Sellke 
\cite{huang2021tight}.

Remarkably, using this idea, \cite{huang2021tight} proves that the algorithmic 
threshold of Theorem \ref{thm:ALG} is tight over the class of Lipschitz algorithms.
\begin{theorem}[\cite{huang2021tight}]
Let $H$ be the spin glass Hamiltonian of Eqs.~\eqref{eq:pspin},
\eqref{eq:GaussianCouplings}. 
If $\bsigma^{\salg}=\bF(H)$ where $\bF: H\mapsto \bF(H)\in [-1,+1]$ is an algorithm that is 
Lipschitz continuous 
(with respect to the Euclidean norm over the 
couplings\footnote{More precisely, recall that the Hamiltonian is parametrized by 
$\bg = (\bG^{(p)}/N^{(p-1)/2})_{2\le p\le p_{\max}}$. Here it is assumed that 
$\|\bF(\bg_1)-\bF(\bg_2)\|_2\le L\|\bg_1-\bg_2\|_2$ or $L$ an $N$-independent constant.}, 
then, almost surely 
\begin{align*}
\lim_{N\to\infty}\frac{1}{N}H(\bsigma^{\salg}) &\ge \ALG:= \sup_{\gamma\in\cuL} \sP(\gamma)\,.
\end{align*}
\end{theorem}

Interestingly, the proof is based on a `geometric' interpretation of the algorithmic threshold 
$\ALG$, that we now sketch. Fix an integer $m$, and a positive definite matrix 
$\bQ=(Q_{ab})_{a,b\le m}$, with $Q_{aa}=1$. Then we can define the energy threshold
$\sE(\bQ)$ to be the minimum energy such that we can find $m$ configurations
$\bsigma^1,\dots,\bsigma^m$ whose normalized inner products (overlaps) are (asymptotically) given by the 
$(Q_{ab})_{a,b\le m}$.
More formally, define
\begin{align*}
\sE_N(\bQ;\delta):= &\min\Big\{E:\;\; \exists \bsigma^a\in\{+1,-1\}^N, a\le m\;\;\mbox{s.t.}\;\;\;\;
\frac{1}{N}H(\bsigma^a)\le  E \;\forall a,\\ 
&\phantom{AAAAAA}\Big|\frac{1}{N}\<\bsigma^a,\bsigma^b\>-
Q_{ab}\Big|\le \delta\;\forall a,b
\Big\}\, ,\\
\sE(\bQ) := &\lim_{\delta\to 0}\lim_{N\to\infty} \sE_N(\bQ;\delta)\, .
\end{align*}
where we assume that $\lim_{N\to\infty} \sE_N(\bQ;\delta)$ exists in probability.

Next we consider a special class of matrices. Let $\mu(q)$ be a distribution function
on the interval $[0,1]$ (i.e., a non-decreasing, right continuous function with $\mu(0)=0$,
$\mu(1)=1$). Denote by $\bQ_m(\mu)$ the $m\times m$ `Parisi matrix' obtained by discretizing $\mu$
(the reader unfamiliar with this construction should consult any of 
\cite{mezard1987spin,MeMo,panchenko2013sherrington}).
With an abuse of notation, we let
$\sE(\mu) := \lim_{m\to\infty} \sE(\bQ_m(\mu))$.
Finally 
\begin{align}
\ALG = \inf\Big\{\,\sE(\mu):\;\; \mu \mbox{ strictly increasing }\Big\}\, .
\end{align}
In physics jargon, $\ALG$ is the smallest energy such that we can construct a full-RSB,
no-overlap gap tree whose leaves are configurations with that energy. 

%
%
\section{Conclusions}
\label{sec:Conclusion}

This report left out many beautiful results and ideas
in the same general research area. 
In this section we mention a few other developments:
the literature is substantial and references are only meant to point the reader towards
some recent works. 

\paragraph {Semidefinite programming hierarchies.} These powerful constructions
allow to define increasingly accurate convex approximations of the non-convex problem of
interest.
 These convex relaxations appear to be limited by the same computational barriers described in the
 previous section. However, the analysis of SDP hyerarchies is significantly more challenging 
 and we are far from a precise or general understanding of the connections
 (see \cite{banks2021local,ivkov2023semidefinite} for recent progress in this direction).
 
\paragraph{Low-degree polynomials.} An intermediate class of algorithm, between Lipschitz 
or message passing algorithms and SDP relaxations is provided by low-degree polynomials.
They could be an ideal ground for understanding connections, see e.g.
\cite{bresler2022algorithmic,montanari2022equivalence,schramm2022computational}.

 \paragraph{Search versus refutation.} In fact, SDP relaxations attempt to solve a harder 
 problem than the style of algorithms discussed here. They try to find an approximate ground state
 $\bsigma^{\salg}$ but also to return a lower bound on the ground state energy 
 which is valid even in the worst case. The latter problem is referred in computer 
 science as `refutation'. Precise results on the possibility/impossibility of refutation are 
 rare \cite{barak2019nearly,deshpande2019threshold,raghavendra2017strongly,kunisky2021tight}.

\paragraph{Reductions.} A different way to provide evidence for hardness of a random optimization
problem ${\sf X}$ is to prove that other seemingly hard random optimization problems 
${\sf Y, Z, \dots}$ reduce to it. Here `reduction' is interpreted in the computer science 
sense. The challenge is that, of course, these reductions must map the 
`natural' probability distribution of ${\sf X}$ to the one of ${\sf Y}$. Despite these challenges,
there has been important progress in proving such reductions recently
\cite{berthet2013complexity,brennan2018reducibility,brennan2020reducibility}.

\paragraph{Sampling.} Sampling from a given high-dimensional probability distribution
is at least as fundamental a computational problem as optimization. In fact,
the two problems are related as suggested by considering the Boltzmann-Gibbs
 distribution $\mu(\bsigma) \propto \exp(-\beta H(\bsigma))$.
However sampling is in general harder because it requires exploring 
configurations at a given energy in a uniform way, and not just finding one. 

Recently there has been interesting progress in developing sampling algorithms 
for the Gibbs measure associated to mean field glassy models \cite{eldan2022spectral,el2022sampling}. However
this line of research is still at its beginnings.

\paragraph{Richer distributions.} Most work surveyed in this report was carried out 
on very simple distributions for the cost function $H$. The prototypical example is
provided by the spin glass model of Eqs.~\eqref{eq:pspin},
\eqref{eq:GaussianCouplings}. It would be important move beyond such examples.
There has been recent progress on large classes of invariant models
\cite{fan2022approximate,dudeja2023universality,wang2022universality}, as 
well as nonlinear random features models \cite{gerace2020generalisation,hu2022universality,montanari2022universality}.
These directions will probably remain active in the future.

\vspace{0.5cm}

In conclusion, the last 40 years have witnessed remarkable progress towards
understanding the fundamental complexity barriers and optimal algorithms to 
solve certain classes of random optimization problems. In a sense, the hypotheses
about connections between algorithms, phase transitions, complexity, and structure 
of the space of solutions that were vaguely formulated in the eighties are now precise, 
and made rigorous in simple yet highly non-trivial models.

It is important to note that this progress was enabled by a truly remarkable interaction between
multiple research communities: physics, computer science, probability theory. 
Results from each community have influenced the others, and ideas have crossed boundaries
multiple times. 

\subsection*{Acknowlegements}

This work was partially supported by the NSF through award DMS-2031883, the Simons Foundation through
Award 814639 for the Collaboration on the Theoretical Foundations of Deep Learning, the NSF
grant CCF-2006489 and the ONR grant N00014-18-1-2729.
%
%

\bibliographystyle{amsalpha}

\newcommand{\etalchar}[1]{$^{#1}$}
\providecommand{\bysame}{\leavevmode\hbox to3em{\hrulefill}\thinspace}
\providecommand{\MR}{\relax\ifhmode\unskip\space\fi MR }
\providecommand{\MRhref}[2]{%
  \href{http://www.ams.org/mathscinet-getitem?mr=#1}{#2}
}
\providecommand{\href}[2]{#2}

\end{document}